\documentclass[namedreferences]{SolarPhysics}
\usepackage[optionalrh]{spr-sola-addons}
\usepackage{epsfig}
\usepackage{graphicx}
\usepackage{color}
\usepackage{rotating}
\usepackage{url}

\begin{document}

\begin{article}

\begin{opening}

\title{LONG-TERM VARIATIONS IN THE GROWTH AND DECAY
RATES OF SUNSPOT GROUPS}

\author{J.\ JAVARAIAH}

\runningauthor{J. JAVARAIAH}
\runningtitle{GROWTH AND DECAY OF SUNSPOT GROUPS}

 \institute{Indian Institute of Astrophysics, Bangalore-560 034, India.\\
email: \url{jj@iiap.res.in}\\
}

\begin{abstract}
 Using  the  combined  Greenwich (1874\,--\,1976) and Solar Optical 
Observatories Network (1977\,--\,2009) data on sunspot groups, we study 
 the long-term variations in  the mean  daily rates of growth 
  and decay   of sunspot groups. We find that 
the minimum and the maximum  values of the  annually averaged daily
 mean growth rates are 
  $\approx$52\%  day$^{-1}$ and $\approx$183\%  day$^{-1}$, respectively,
 whereas the corresponding values of the annually averaged daily mean
 decay rates
 are $\approx$21\% day$^{-1}$ and  $\approx$44\%  day$^{-1}$, respectively. The average value (over the
 period 1874\,--\,2009) of the  growth rate    is about 70\% more than that of 
  the decay rate. The growth  and the decay rates vary by about 35\%
 and 13\%, respectively,  on a 60-year 
time-scale. From the beginning of Cycle~23  
the growth rate is substantially decreased and   near the end  
(2007\,--\,2008) 
the growth  rate is lowest in the past about 100 years. 
In the extended part  of the
 declining phase of this cycle, the decay 
rate steeply increased  and it is largest in the
 beginning of the current Cycle~24. 
These unusual properties of the growth and the decay rates  during Cycle~23
  may be related to cause of the  very
  long declining phase of this cycle  with the 
 unusually  deep and prolonged current minimum.
A $\approx$ 11-year periodicity in the growth and the decay rates 
is  found to be highly latitude and time 
dependent and seems to exist mainly in  the $0^\circ - 10^\circ$ latitude 
interval of the southern hemisphere. 
The strength of the known  approximate
 33\,--\,44 year modulation in the solar activity seems to be related to   
 the  north-south asymmetry in the growth rate.
Decreasing and increasing   trends in the growth and the decay rates
 indicate that the next 2\,--\,3 solar cycles will be relatively weak. 
\end{abstract}
\end{opening}

\section{Introduction}
Magnetic flux,  in the form of large  flux tubes, emerges 
 to the surface--presumably 
  from near the base of the convection zone (where the dynamo process 
is believed  to be taking place)--and  
 responsible for sunspots and 
other solar active phenomena (see Rosner and Weiss, 1992; Gough, 2010).
 A sunspot lasts from a few hours to several weeks.
 The typical sizes of sunspots range from
 ~10 msh (millionth of the solar hemisphere $\approx 3\times10^6$ km$^2$) 
to $10^3$ msh.
Although individual sunspots
are common,  the majority of sunspots are parts of  groups.
 Spot groups are often large
 and complex.
 The daily area of a spot (or spot group) is one of the
most important
parameters used to describe the spot (or spot group) development. 
The area of a spot (or spot group) is closely connected with the
magnetic flux of the spot (or spot group)
(130 msh area
 $\approx\ 10^{22}$ Mx magnetic flux, $e.g.$, see Wang  and Sheeley, 1989).
That is, the development of the
spot (or spot group) area  reflects the development of
the solar magnetic field.
 Therefore,    
the  increase/decrease  in the  areas of spots or 
 spot groups, $i.e.$, the  growth/decay of spots or spot groups, 
 can affect
significantly the strength, configuration and topology of the magnetic structure in
the solar atmosphere. Hence, 
  the studies of  growth and decay of 
sunspots or sunspot groups are important 
 for understanding configuration  and topology of the 
magnetic structure on the solar surface,  the solar variability and  
the underlying 
 mechanism of it.
Several such studies have been made and many characteristics of the growth and 
decay of the spot groups are found (see Lustig and W\"ohl, 1995; 
Hathaway and Choudhary, 2008). 

Howard (1992a,1992b) analyzed Mt. Wilson sunspot and sunspot group  data 
during 1917\,--\,1985 and determined 
many properties of the day-to-day changes 
in the sunspot umbral areas (spot growth/decay). 
Howard (1992a) also studied the variations in  the annual averages of
 the umbral area 
increases, but no 
 systematic variations  are found.
In the present study we analyze a large data set of sunspot groups and 
 attempt to detect  the long-term 
variations  in  daily rates of growth and decay of sunspot groups. 

In the next section we  describe the methodology and the data analysis. 
In Section~3 we  present  the results. 
In Section~4 we  draw 
conclusions and
briefly discuss the implications of them on the long-term solar variability. 

\section{Methodology  and data analysis}
 Here we have used the combined  Greenwich  
(1874\,--\,1976) and  Solar Optical Observation 
Network (SOON) (1977\,--\,2009) sunspot group
data, which are    
 taken  from  David Hathaway's  website 
{\tt http://solarscience.msfc.nasa.gov/greenwch.shtml}.
These data  
included the observation time (the Greenwich data contain the date with the
fraction of the day, in  the SOON data  
 the fraction  is rounded to 0.5 day), 
heliographic latitude ($\phi$) and  
longitude ($L$), central meridian distance (CMD), and 
corrected umbra and whole-spot areas (in msh), etc., of the spot 
groups for each day of observation. 
The 
positions of the groups  are geometrical  positions of the 
centers of the groups.

The Greenwich data have been compiled from the majority of the white
light photographs which were secured at the Royal Greenwich Observatory
and at the Royal Observatory, Cape of Good Hope. The gaps in their
observations were filled with photographs from  other observatories,
 including the Kodaikanal Observatory, India.
The SOON data  included measurements made   by the 
United States Air Force (USAF) from 
the sunspot drawings of a network of the observatories
 that has included telescopes
in Boulder, Colorado; Hawaii; etc.
 David Hathaway  scrutinized 
the Greenwich and SOON data and produced a reliable 
continuous data series
from 1874 up to date (also see Hathaway and Choudhary, 2008).
In case of SOON data, we increased the area by a factor of 1.4.
 David Hathaway found this correction was necessary to
 have a combined homogeneous
 Greenwich and SOON
data (see the aforementioned website of David Hathaway.)

The method of analysis is similar to that in Howard (1992a).
 Howard used
 daily umbral areas of the spots
 measured in Mt. Wilson Observatory during the years 1917\,--\,1985.   
We have used the corrected daily whole-spot areas  
(umbral value + penumbral value) of spot groups ($A$).  
A spot group is included when the  observations of it are available for two 
or more consecutive days. 
The daily rate of change of the area ($\frac{\Delta{A}}{\Delta{t}}$) of
  a spot  group is computed using the differences between 
the epochs of its observation 
in 
consecutive days and between the   corrected whole-spot areas 
 of the spot group at these epochs. That is, 
 $$\frac{\Delta{A}}{\Delta{t}} = \frac{A_{\rm n}-A_{\rm n-1}}{t_{\rm n}-t_{\rm n-1}}\ , \eqno(1) $$
\noindent where $t$ is the epoch of observation  during 
the life time ($T$) of a spot group and ${\rm n} =$ 2, 3,...., $T$.   
Positive and negative values of $\frac{\Delta{A}}{\Delta{t}}$  correspond to
 the 
daily rates of  growth ($G$)  and decay ($D$)  of a spot group,  respectively.  
The percentage of growth (\%G) and decay (\%D) are calculated as  
$(G\times 100)/A_{\rm n-1}$ and  $(D\times 100)/A_{\rm n-1}$, respectively.
The mean values ($\overline {\%G}$ and $\overline {\%D}$)  of $\%G$ and $\%D$
 in a given time interval are calculated as follows:
$$\overline {\%G} =\frac{1}{\rm k} \Sigma \%G_i\ \ {\rm and}\ \  
\overline {\%D} =\frac{1}{\rm m} \Sigma \%D_j\ ,  \eqno(2) $$
\noindent where $i=$ 1, 2,....,k  and $j=$ 1, 2,....,m;    
 k and m   are 
the number of data points of $\%G$ and $\%D$, respectively,  
in the interval. 
It should be noted that obviously more contributions to 
$\overline {\%G}$ and $\overline {\%D}$  are 
coming from  the spot groups 
 before and after reaching their  
maximum areas, respectively.

The data on the  recurrent and the non-recurrent spot groups are combined. 
The first and the subsequent disc passages of the recurrent spot groups are 
treated as independent groups. 
Therefore, the   life time    of a spot group is
$\le$ 12 days.   
As in the case of our  earlier papers (Javaraiah 2010, and references therein), 
here  we have taken the following precaution which 
 reduces
substantially the
 uncertainties in the derived results (Ward, 1966;
Javaraiah and Gokhale, 1995):
We have excluded the
data corresponding to the $|CMD| > 75^\circ$ on any day of the spot group
life span. This  reduced  the error  due to the  
foreshortening effect.
Further, we  excluded the data corresponding to
 the `abnormal'  motions, $e.g.$  displacements
exceeding $3^\circ$ day$^{-1}$
  in the longitude or  $2^\circ$ day$^{-1}$ in the latitude. This reduces 
 uncertainties in $\overline {\%G}$ and $\overline {\%D}$
  removing those which were incorrectly identified.
For the sake of  readers convenience,  in Table~1 and Figure~1 we have  
demonstrated the determination of  $\%G$ and $\%D$ for one spot group: 
   NOAA/USAF  spot group no. 2288,  
 observed during the days 18.5\,--\,29.5 of 
February 1980 
(Note: Greenwich sunspot group number 
through 1976; NOAA/USAF group number after 1976). 
 If there is only this spot group in the given 
time interval (say in one year, 1980), then   
  $\overline {\%G}$ is 105.7,  determined 
from the five values of $\%G$ at the epochs 
  19.5\,--\,20.5, 20.5\,--\,21.5, 21.5\,--\,22.5,  
24.5\,--\,25.5, and 26.5\,--\,27.5  
during the life time of this spot group, 
 and  $\overline {\%D}$ is 22.4,  determined  
from the three values of $\%D$ at the epochs 22.5\,--\,23.5, 23.5\,--\,24.5, 
and  25.5\,--\,26.5.
The standard errors of these mean values are $\sigma_g/\sqrt(k) = 63.9$ and 
$\sigma_d/\sqrt(m) =17.3$, where $\sigma_g$ and $\sigma_d$ are the standard 
deviations correspond to  $\overline{\%G}$ and $\overline{\%D}$,
 respectively.   
The value of $\%G$ at the epoch  18.5\,--\,19.5 is not considered because
$\frac{\Delta \phi}{\Delta t} > 2^\circ$ day$^{-1}$ 
 and the value of $\%D$ at the epoch 27.5\,--\,28.5 is not considered because  
$\frac{\Delta L}{\Delta t} >3^\circ$ day$^{-1}$ and   $\frac{\Delta \phi}{\Delta t} >2^\circ$ day$^{-1}$.  
The value at the epoch 28.5\,--\,29.5 is
not considered because the $CMD$ at the epoch 29.5 is $> 75^\circ$. 

\begin{table*}
   \caption{The data of the spot group,  NOAA/USAF spot group no. 2288, 
  observed during the days 18.5\,--\,29.5 of 
February 1980. A positive and a negative value of $\frac{\Delta A}{\Delta t}$ 
(differences of $A$ on  consecutive days) 
 represent  
the growth rate ($G$) and decay rate ($D$) 
 respectively.}
\begin{tabular}{lccccccc}
\hline
  \noalign{\smallskip}
$t_{\rm n}$ &$L$&$\phi$&$CMD$&$A$&$\frac{\Delta A}{\Delta t}$& $\%G\ {\rm or}\ \%D$\\
\hline
  \noalign{\smallskip}
   18.5&    319.5&  13.0& -67.0&   10&     10&  100.0\\
   19.5&    317.4&  10.0& -56.0&   20&     70&  350.0\\
   20.5&    319.2&   9.0& -41.0&   90&     60&   66.7\\
   21.5&    320.0&  10.0& -27.0&  150&    150&  100.0\\
   22.5&    319.9&  11.0& -14.0&  300&    -20&   -6.7\\
   23.5&    320.7&  10.0&   0.0&  280&    -10&   -3.5\\
   24.5&    320.5&  10.0&  13.0&  270&     10&    3.7\\
   25.5&    320.4&  10.0&  26.0&  280&   -160&  -57.1\\
   26.5&    320.2&  10.0&  39.0&  120&     10&    8.3\\
   27.5&    321.0&  11.0&  53.0&  130&    -90&  -69.2\\
   28.5&    325.9&  10.0&  71.0&   40&     0&    0.0\\
   29.5$^{\mathrm{a}}$&    327.7&   8.0&  86.0&   40&      & \\
\hline
  \noalign{\smallskip}
\end{tabular}

$^{\mathrm{a}}$ indicates that this day data are not used here due to 
$CMD >75^\circ$.
\end{table*}

\begin{figure}
\centerline{\includegraphics[width=\textwidth]{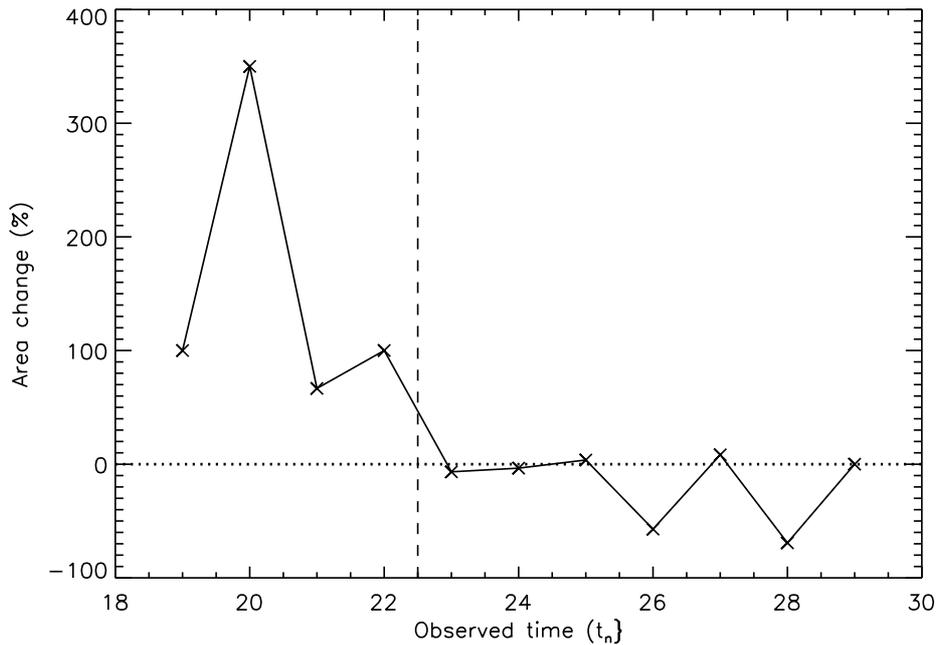}}
\caption{Plot of the percentage area change in the consecutive days  
during the life time  
 of the spot group, NOAA/USAF spot group no. is 2288, 
 observed during the days 
 18.5\,--\,28.5 of February 1980, {\it versus} time (middle epoch of a given 
consecutive day). The vertical dashed-line represents 
the epoch of the spot group maximum area. The positive and negative values 
 correspond to the percentage growth day$^{-1}$ ($\%G$) and percentage 
decay day$^{-1}$ ($\%D$), respectively. The horizontal dotted-line is drawn
at $\%G = \%D =0$.}  
\end{figure}

Hathaway and Choudhary (2008)   analyzed the combined Greenwich 
and SOON data and 
 found that the  decay rates of the spot groups of   Cycles~21\,--\,23 are 
much larger than those of  Cycles~12\,--\,20.
 We  noticed 
that some daily data records, particularly in the data 
during  recent cycles ($i.e.$, in the SOON dataset), 
 contain the zeros for the whole-spot group
 area. We excluded all those unrealistic data records, $i.e.$,
  we have excluded 
 the  data correspond to the whole-spot area equal to zero in any day during 
the life time  of the spot group, because we find that presence of these 
unrealistic data yields large values for the derived decay rates.  

We  determined  the variations in the
 mean percentage growth and decay rates 
($\overline {\%G}$ and $\overline {\%D}$) of the  
  spot groups  in the 
 whole disk, in the   
 northern 
and  the  
southern  hemispheres, 
 and also  in the separate  
$10^\circ$ latitude intervals.  
The variations in $\overline {\%G}$ and $\overline {\%D}$ of
  the spot groups  in the whole disk 
are determined from 
   the yearly data and for the spot groups in
 a whole hemisphere the variations are determined
 by binning  the data into  3- and 4-year moving time intervals  
 (MTIs) that are successively shifted by one year during the period 1974\,--\,2009, for the sake of better statistics. 
In case of the separate $10^\circ$ latitude intervals, we  have used only   
4-year MTIs because in a shorter than  4-year interval 
 the data are found to be inadequate and the error bars are very large 
 to plot the results, particularly during the cycles minima. 
The yearly as well as 3- and 4-MTIs time series of  
 $\overline {\%G}$ and $\overline {\%D}$
 have been corrected by replacing  
 those values of  $\overline {\%G}$ and $\overline {\%D}$
whose standard errors exceeded the 2.6 times
(correspond to 99\%  confidence level) the corresponding median values with 
the average of the corresponding values and their respective two neighbors
 (in case of the  beginnings of the 
time series
  it is the average of 
 the values in the intervals 1 and 2 and in the endings    
 it is the average of the  values in the intervals $N-1$ and $N$, where 
N is the size of the series).
We  determined the periodicities in $\overline {\%G}$ and $\overline {\%D}$ 
 from the  fast
Fourier transform (FFT) power spectrum analysis of the corrected
time series.  The values of the
periodicities are determined
 from the maximum entropy method (MEM).
The time-dependencies in the periodicities in
  $\overline {\%G}$ and $\overline {\%D}$  are checked 
 using  Morlet wavelet analysis. 
 The MEM and the wavelet analyses of
  $\overline {\%G}$ and $\overline {\%D}$ were 
carried out in a similar way  
as in
 the  analyses of the mean 
 meridional motions of the spot groups by Javaraiah (2010) and are  
 briefly
 described  below.   

 The lengths of the  time series are inadequate to measure precisely 
the values of $\ge$ 11-year periodicities in
 $\overline {\%G}$ and $\overline {\%D}$ from the FFT analysis. Hence,   
 the uncertainties in the longer  
 periodicities determined here  from the FFT analysis are large.
 A different approach for determining the  periodicities 
in a short time series with a higher accuracy is to compute
the power spectrum using MEM.
MEM analysis is a parametric modeling approach to the estimation of the power 
spectrum of a time series. The method is data adaptive being used upon an 
autoregressive modeling process. 
An important step in this method is the optimum selection 
of the order $M$ of the autoregressive process,
 which is the number of immediately previous points that have been used 
in the calculation of a new point. 
If $M$ is chosen too low the spectrum is over-smoothed and the high resolution 
potential is lost. If $M$ is chosen too high, frequency shifting and 
spontaneous splitting 
 of the spectral peaks occur.  
The MEM code which we have used here takes the values for $M$ 
 in the range (N/3, N/2) (Ulrych and Bishop, 1975)   or 2N/$\ln$(2N)
(Berryman, 1978).  
In order to find the correct values of the periodicities 
found in the FFT power spectrum, 
 we have computed  
MEM power spectra choosing  various 
values for   M in the range (N/3, N/2) and  2N/$\ln$(2N). 
We find that ${\rm M} = {\rm N/3}$ is  suitable in the present MEM 
analysis,  $i.e.$, in the derived spectra the peaks are considerably  sharp 
and well separated.

Wavelet analysis   provides both time-domain information and 
frequency-domain information simultaneously. 
We have used the wavelet IDL code 
provided by Ch. Torrence and G. P. Compo as described in 
  Torrence and Compo (1998).
Morlet wavelet consists of a complex exponential modulated by a 
Gaussian,  
$exp(\omega_0t/s) exp(-t^2/(2s^2)$, where $t$ is the time, $s$ is the
 wavelet scale, and $\omega_0$ is a non-dimensional frequency. 
A good temporal resolution is required to localize the power maxima in time
and a good frequency resolution is required to determine the 
corresponding frequencies. 
A narrow (in time) function will have good time resolution but poor frequency
 resolution, while a broad function will have poor time resolution, yet good
 frequency resolution. 
For $\omega_0 = 6$ (used here), there are approximately three oscillations
 within the Gaussian envelope. 
According to   Torrence and 
Compo (1998)
it is convenient to write the scales as fractional powers of two:
$ s_j = s_0 2^{j\delta j}, \ \ j = 0, 1, ..., J\  {\rm and}\ J = \delta j^{-1} 
log_2(N\delta t/s_0)$, where $s_0$ is the smallest resolvable scale 
and $J$ determines the largest scale. 
The $s_0$ should be chosen so that the equivalent Fourier 
period is approximately $2\delta t$. The choice of a sufficiently small 
$\delta j$ depends on the width in spectral-space of the 
wavelet function.
 For the Morlet wavelet,  $\delta j =  0.5$  is 
the largest value. 
The wavelet scale $s$ is almost identical 
to the corresponding Fourier period, i.e., 
 the Morlet wavelet with $\omega_0 = 6$ gives $\lambda = 1.03 s$, 
where $\lambda$ is the Fourier period.  
Regions where edge effects become important, because of the finite length of the time series, 
are labeled as cone of influence (COI). The time series is 
padded  with sufficient zeroes to bring the total length $N$ 
up to the next power of two,  limiting the edge effects 
and speeding up the Fourier transform
 (for more details see Torrence and Compo, 1998).
In the present analysis   $N = 136$, $\delta t = 1$ year, $s_0 = 2\delta t$, 
$\delta j =0.105$ and $ J = 58$. The  wavelet spectra of 
$\overline{\%G}$ and $\overline{\%D}$ determined from the corrected
 annual data  have  reasonably good resolutions both  in time and 
frequency.

\section{Results}
Figure~2 shows the  variations in $\overline {\%G}$ and $\overline {\%D}$ 
($cf.$, Equation~2) 
of the sunspot groups in the whole Sun,  
  determined from the
yearly spot group  data during 1874\,--\,2009.
Figure~3 shows the variations in $\overline {\%G}$ and $\overline {\%D}$ 
of the spot groups  in the northern and the southern hemispheres, 
    determined from
the data in 4-year MTIs (3-year MTIs series is not shown because it is
 found to be almost the same as that of 4-year MTIs series).
Figures~4 and 5 show  the variations in $\overline {\%G}$ and $\overline {\%D}$
 of the spot groups
in different 10$^\circ$ latitude intervals determined from the
data in 4-year MTIs.
To study the solar cycle
variations in  $\overline {\%G}$ and $\overline {\%D}$,   
in all of these figures we have also shown the variations in the
sunspot activity. (Note: as per the definition of $D$,  
  $\overline {\%D}$ has negative sign. For the sake of convenience we have 
plotted its  absolute values.)

\begin{figure}
\centerline{\includegraphics[width=\textwidth]{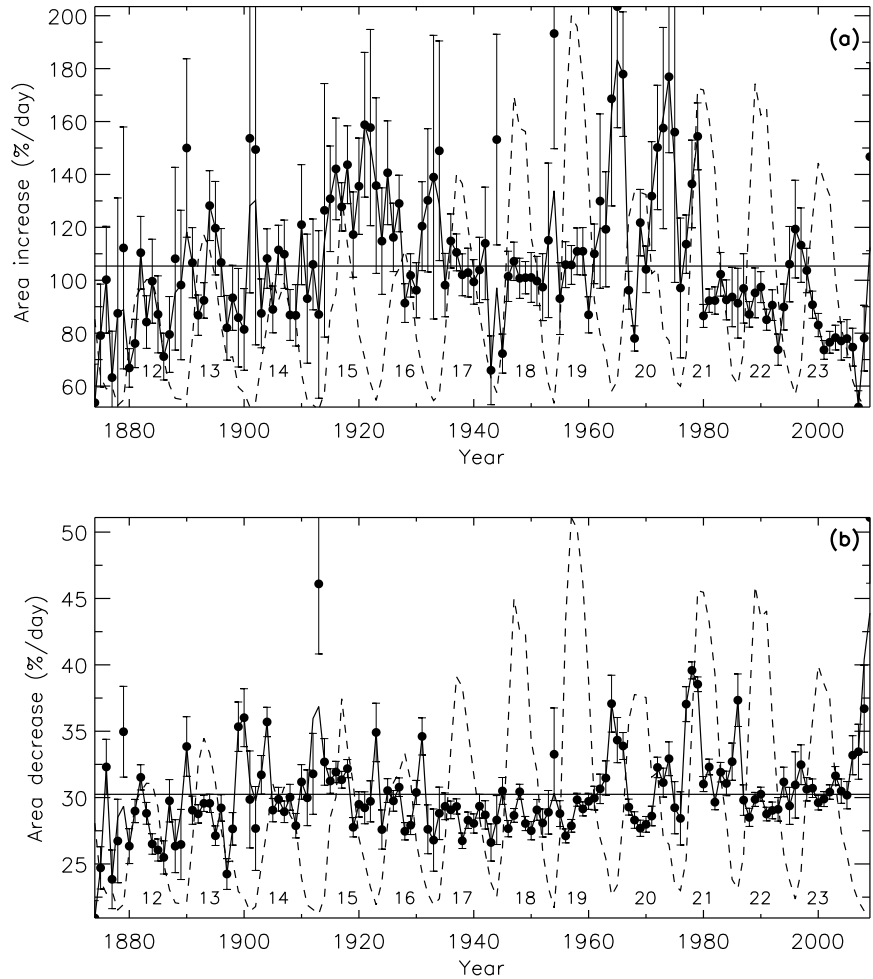}}
\caption{Plots of annual values of the mean percentage rates of the increase
($\overline{\%G}$, upper panel)
and the decrease ($\overline{\%D}$, lower panel) in the
daily area
of the sunspot groups ($cf.$, Equation~(2)) in the whole Sun {\it versus} time,
  during the period 1874\,--\,2009.
The unconnected points represent  the  values
which have a large uncertainty, $i.e.$ standard error
$> 2.6$. The dashed curve represents the
  annual variation
in sunspot
activity during 1874\,--\,2009.
The  Waldmeier cycle number is specified  near the maximum
epoch of each cycle.}
\end{figure}

\begin{figure}
\centerline{\includegraphics[width=\textwidth]{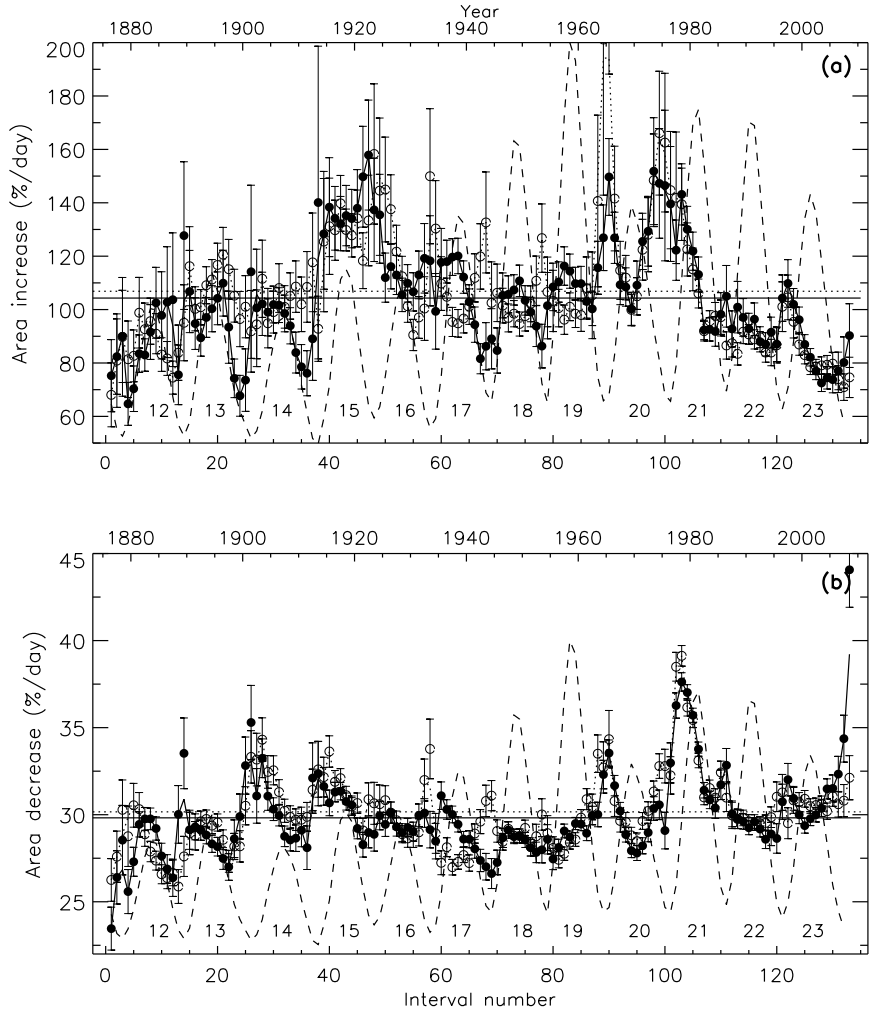}}
\caption{Variations in the mean percentage  daily rates of increase 
($\overline{\%G}$, upper panel) and decrease ($\overline{\%D}$, lower panel) 
in the area of the spot groups
 in the  northern
hemisphere (filled circle-solid curve) and
the southern hemisphere (open circle-dotted curve) 
 determined from 4-year MTIs, 1874\,--\,1877, 1875\,--\,1878, ...,
 2006\,--\,2009. 
The unconnected points represent  the  values 
which have a large uncertainty, $i.e.$ standard error $> 2.6$.
The
 dashed curve represents the
  variation of the sunspot number
smoothed by taking 4-year running average.}
\end{figure}

\begin{figure}
\centerline{\includegraphics[width=\textwidth]{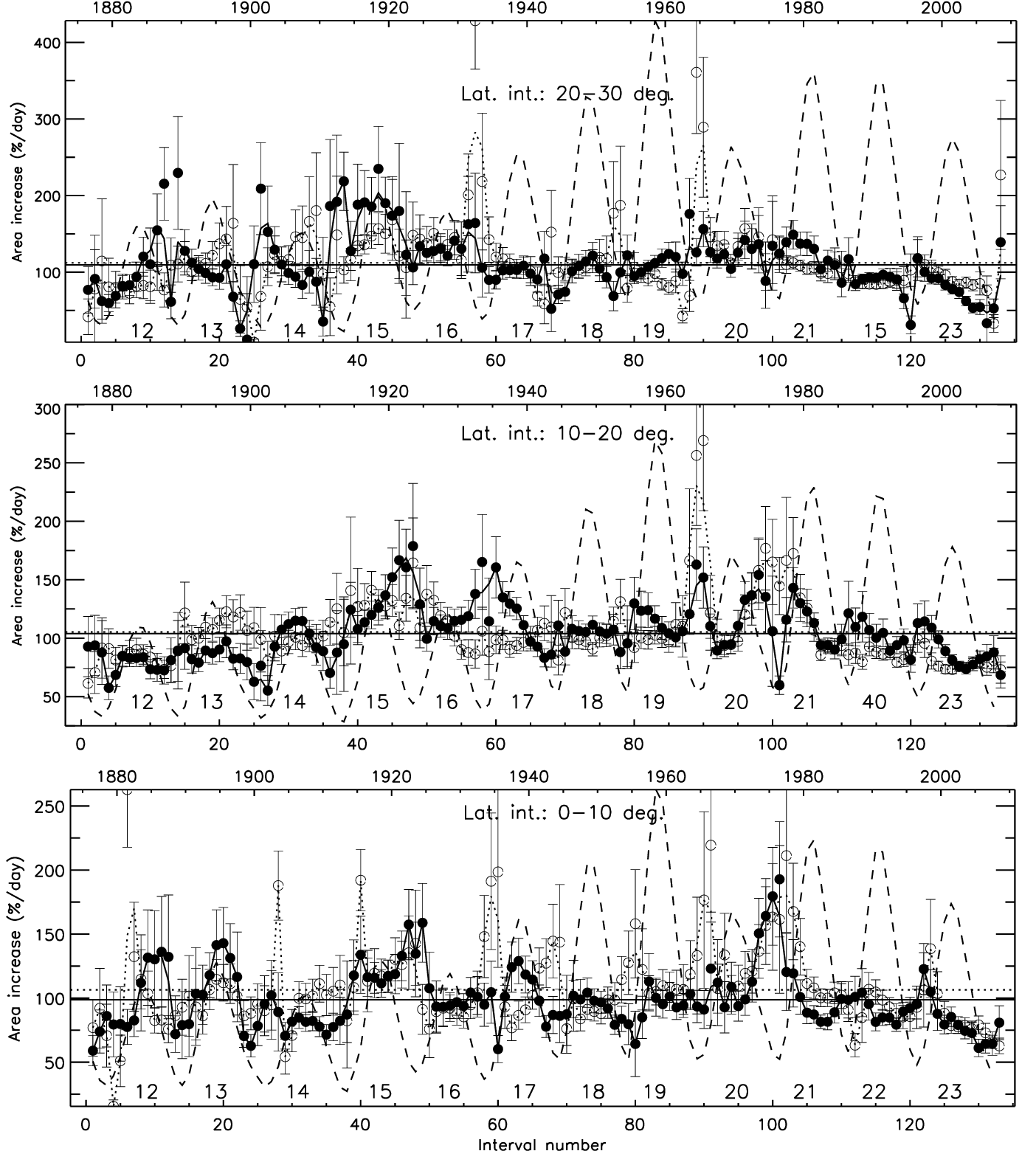}}
\caption{Same as Fig.~3a, but
determined
 separately for different 10$^\circ$ latitude intervals.}
\end{figure}

\begin{figure}
\centerline{\includegraphics[width=\textwidth]{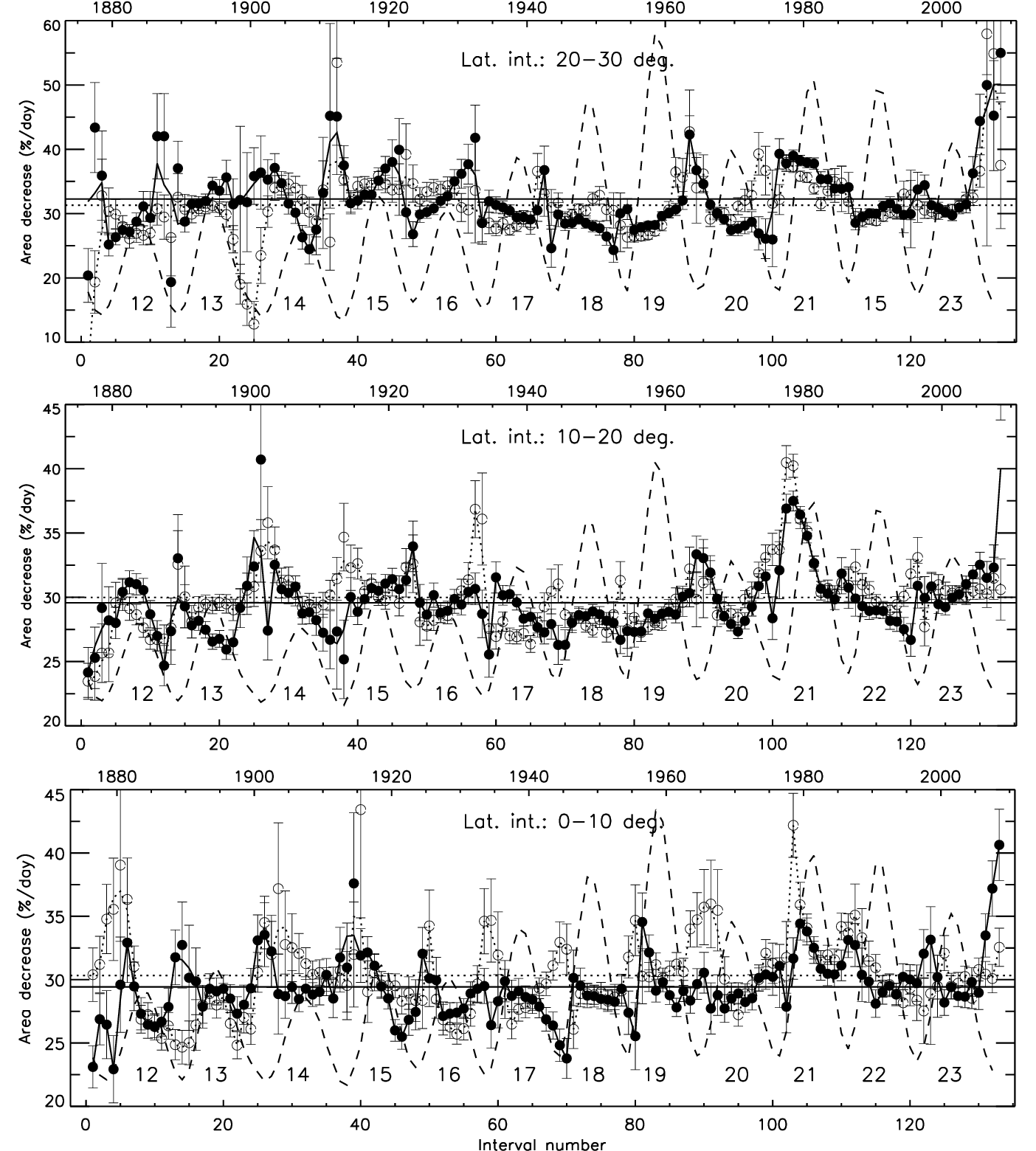}}
\caption{Same as  Fig.~3b, but
determined
 separately for different 10$^\circ$ latitude intervals.}
\end{figure}

\begin{figure}
\centerline{\includegraphics[width=\textwidth]{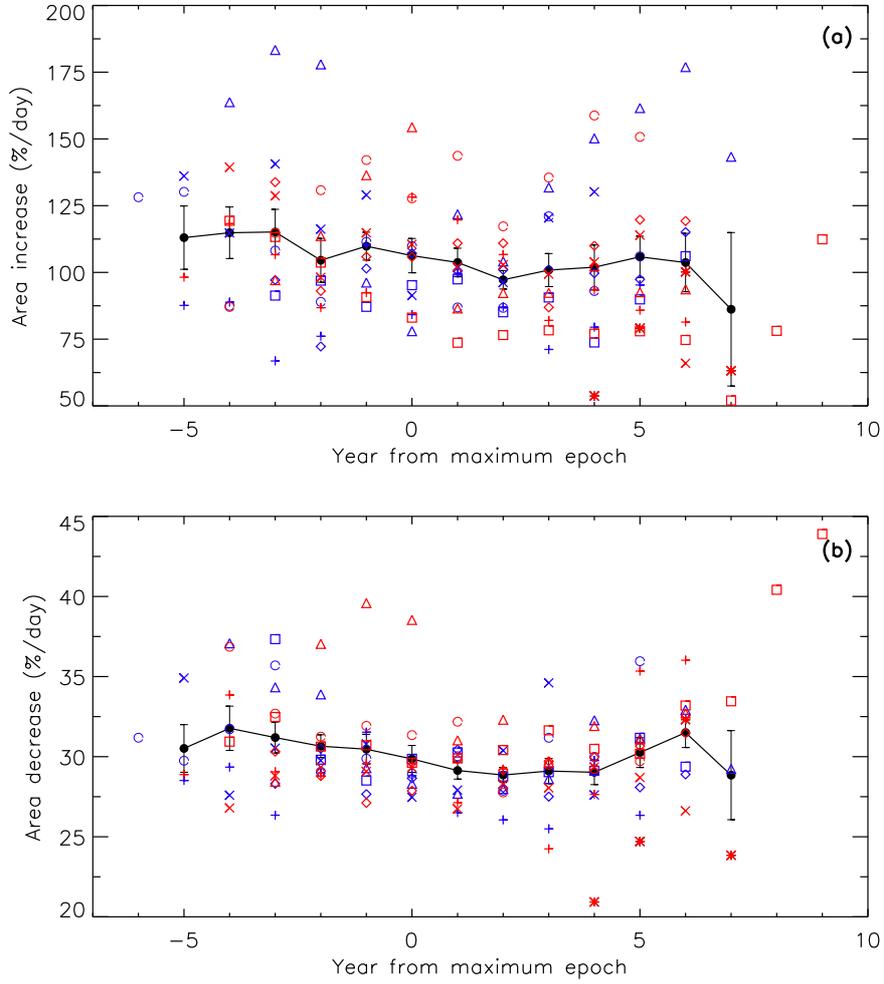}}
\caption{Plots of the yearly mean values (the corrected 
data points which are connected by continuous curves  in Figure~2) of 
$\overline {\%G}$ (upper panel) and $\overline {\%D}$ (lower panel) 
{\it versus} the year from maximum epoch of the solar cycle. 
The red and blue colors are used for odd- and even-numbered 
cycles, respectively.  The different symbols are used for 
different cycles (numbers are given in brackets): asterisks (11), 
pluses (12 and 13), 
open-circles (14 and 15), crosses (16 and 17), diamonds (18 and 19), 
triangles (20 and 21), and squares (22 and 23). 
 The filled circle-continuous curve represents
 the mean solar cycle variation determined from the yearly mean values.
 The error bar represents  the standard error. 
There is only one 
data point  at  years -6 (begin of Cycle~14), 8 (end of Cycle~23) and 
9 (beginning of Cycle 24).}     
\end{figure}

\begin{figure}
\centerline{\includegraphics[width=\textwidth]{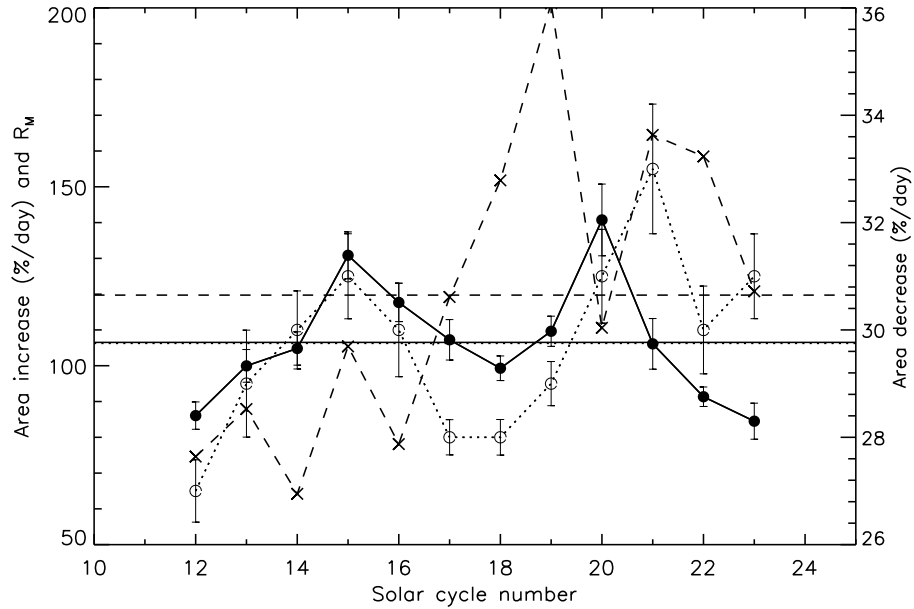}}
\caption{Cycle-to-cycle modulations in
 $\overline {\%G}$  (filled circle-solid curve) and $\overline {\%D}$ 
(open circle-dotted curve) determined by averaging (over a whole 
solar cycle) the  corrected annual values   
shown in  Figure~2.
 The error bar represents the standard error. 
The cross-dashed curve represents the the maximum amplitude ($R_{\rm M}$, the 
largest smoothed monthly sunspot numbers) of the solar cycles.
 The continuous 
and dotted horizontal lines (they overlap) represent the average (over all
 12 cycles) 
values 106.52 and 29.75 of  $\overline {\%G}$ and $\overline {\%D}$, 
respectively. The dashed line represents the mean value (119.74) of 
 $R_{\rm M}$.}
\end{figure}

\begin{figure}
\centerline{\includegraphics[width=\textwidth]{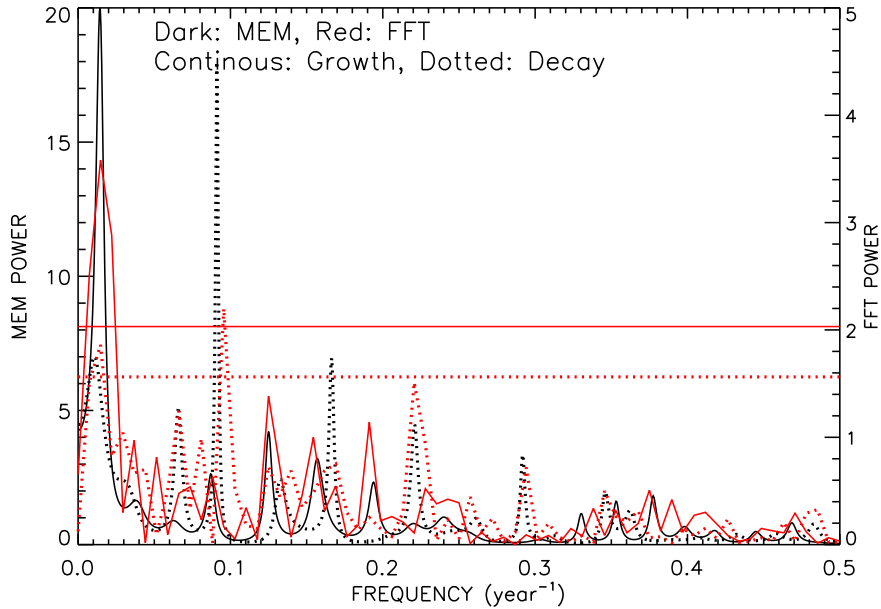}}
\caption{FFT and MEM  power spectra 
of 
the corrected  yearly data of  $\overline {\%G}$ and $\overline {\%D}$  
(whole sphere data) 
shown in  Figure~2. The continuous 
and dotted horizontal lines represent the  95\% confidence levels of the 
peaks in the corresponding FFT spectra.}
\end{figure}

\begin{figure}
\centerline{\includegraphics[width=\textwidth]{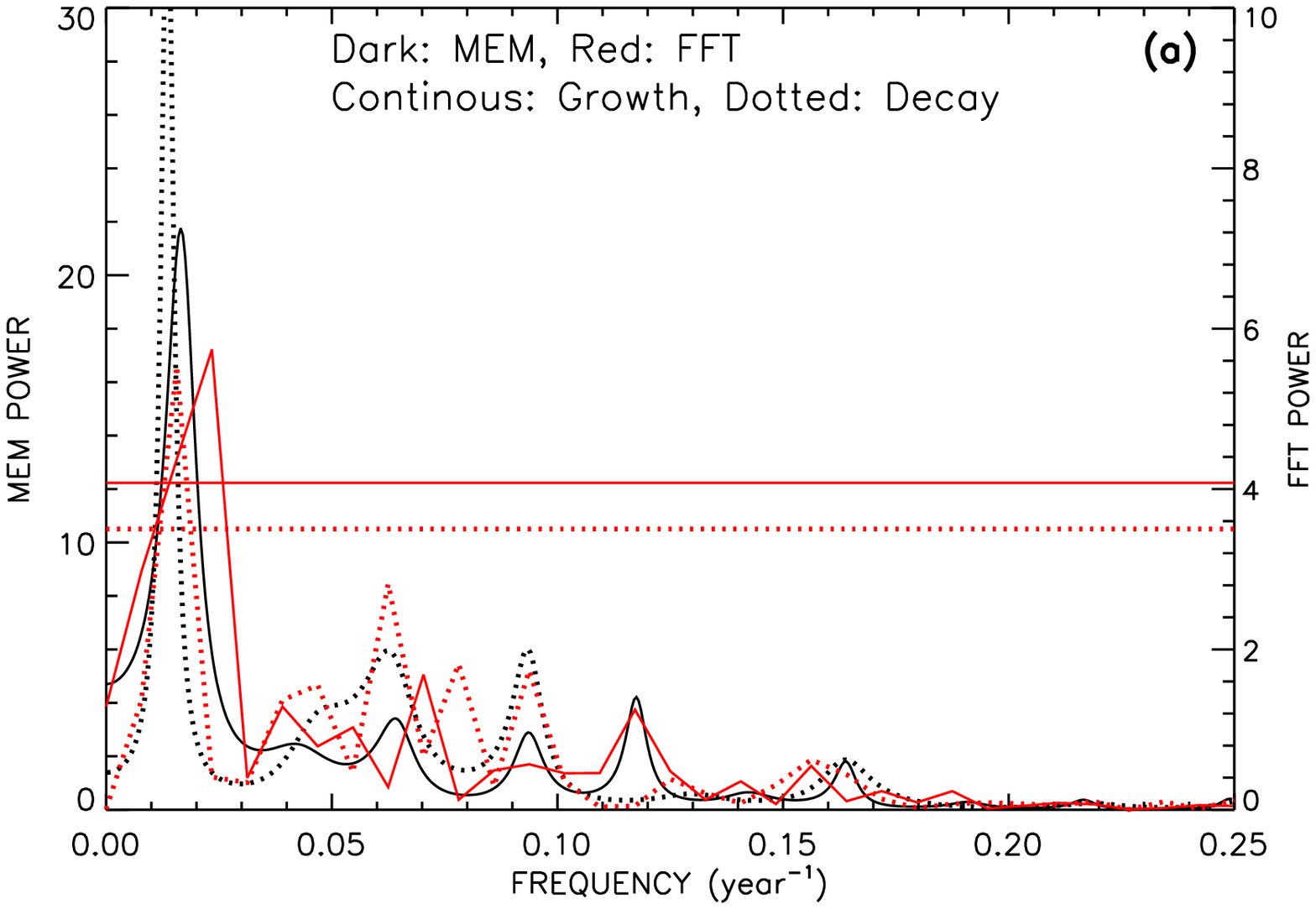}}
\centerline{\includegraphics[width=\textwidth]{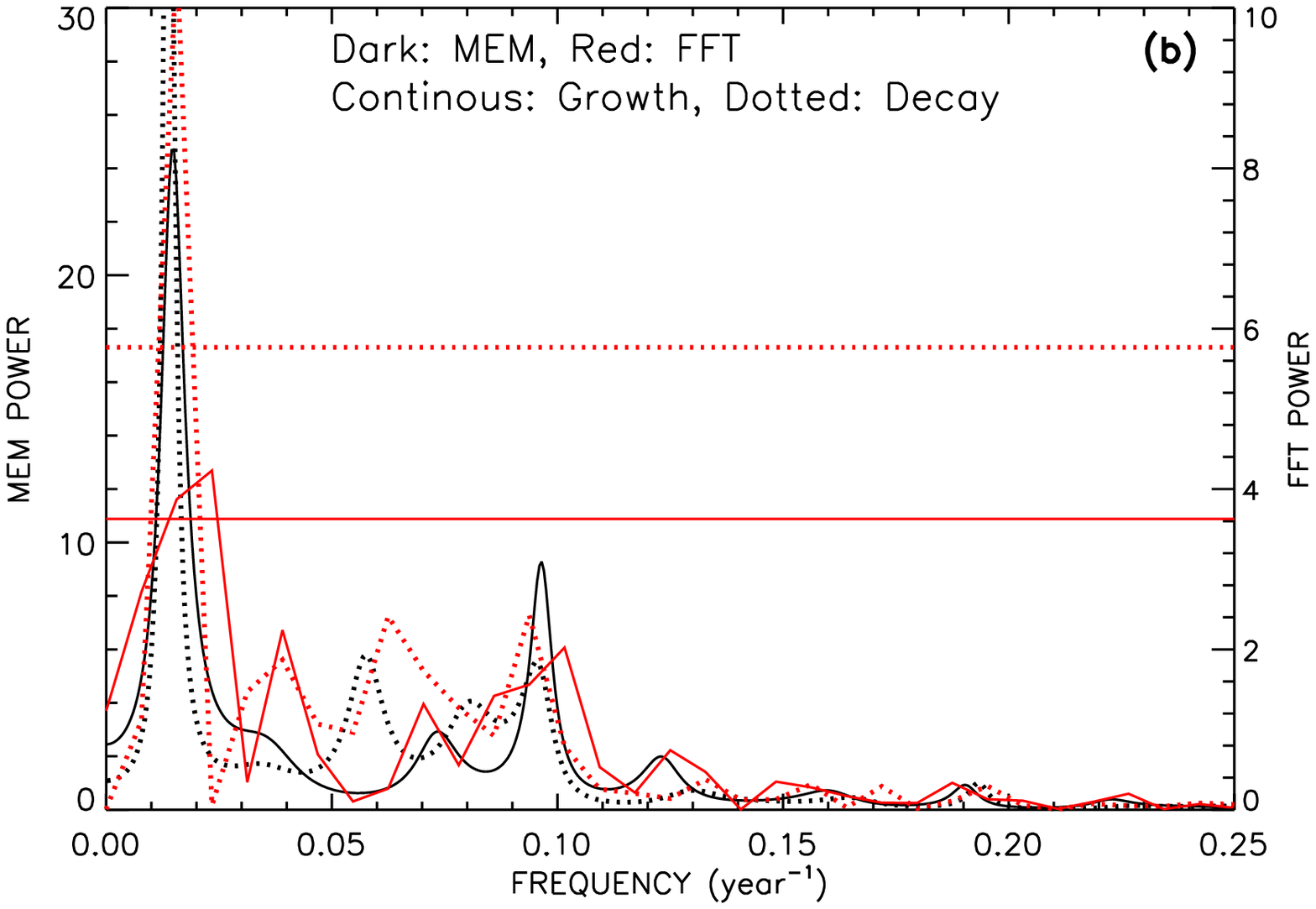}}
\caption{FFT and MEM power spectra 
of $\overline {\%G}$ and $\overline {\%D}$  determined  
(a) from the whole northern hemisphere data and 
(b) from the whole southern hemisphere data in   
   4-year MTIs  shown in  Figure~3.}
\end{figure}

\begin{figure}
\centerline{\includegraphics[width=\textwidth]{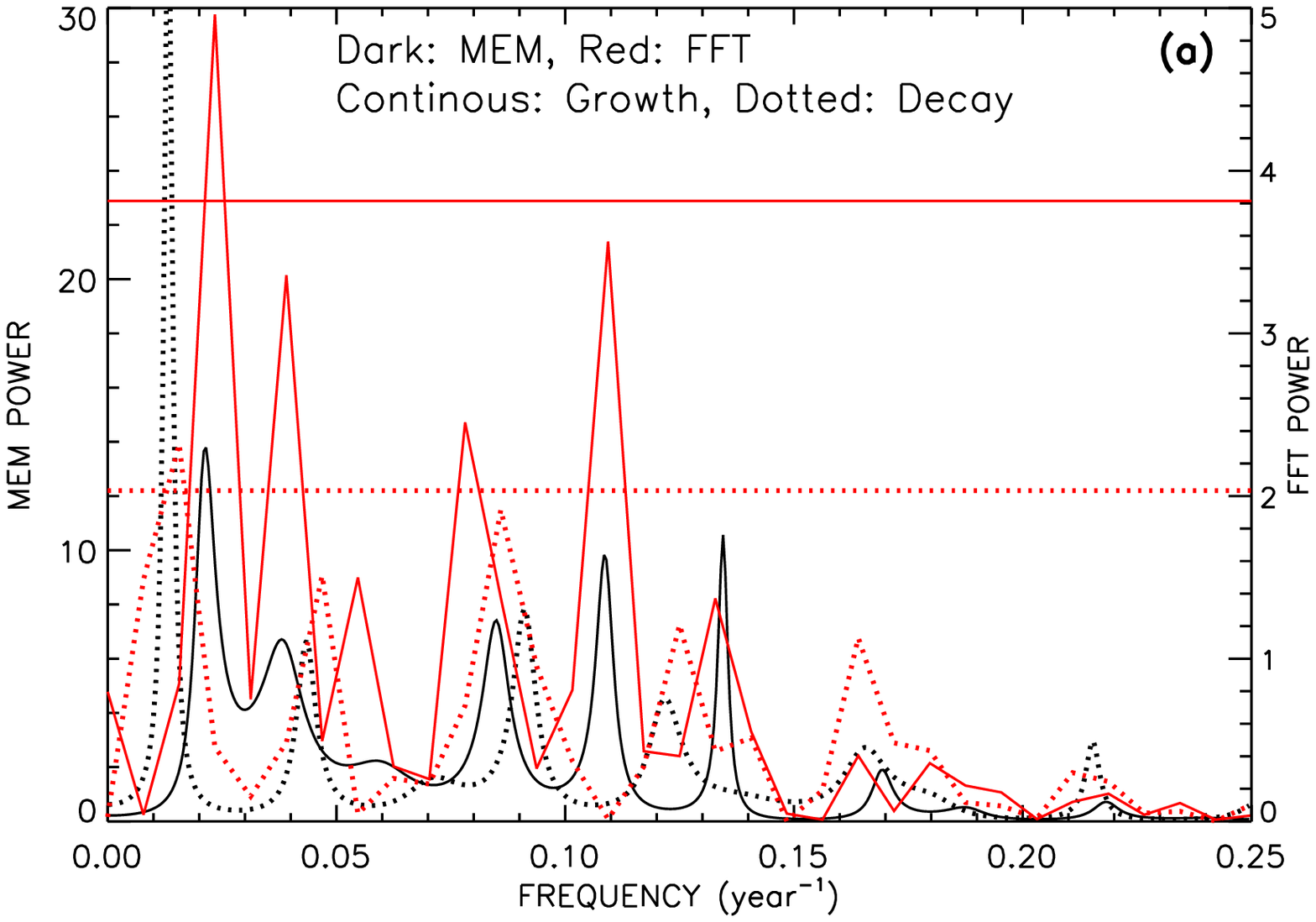}}
\centerline{\includegraphics[width=\textwidth]{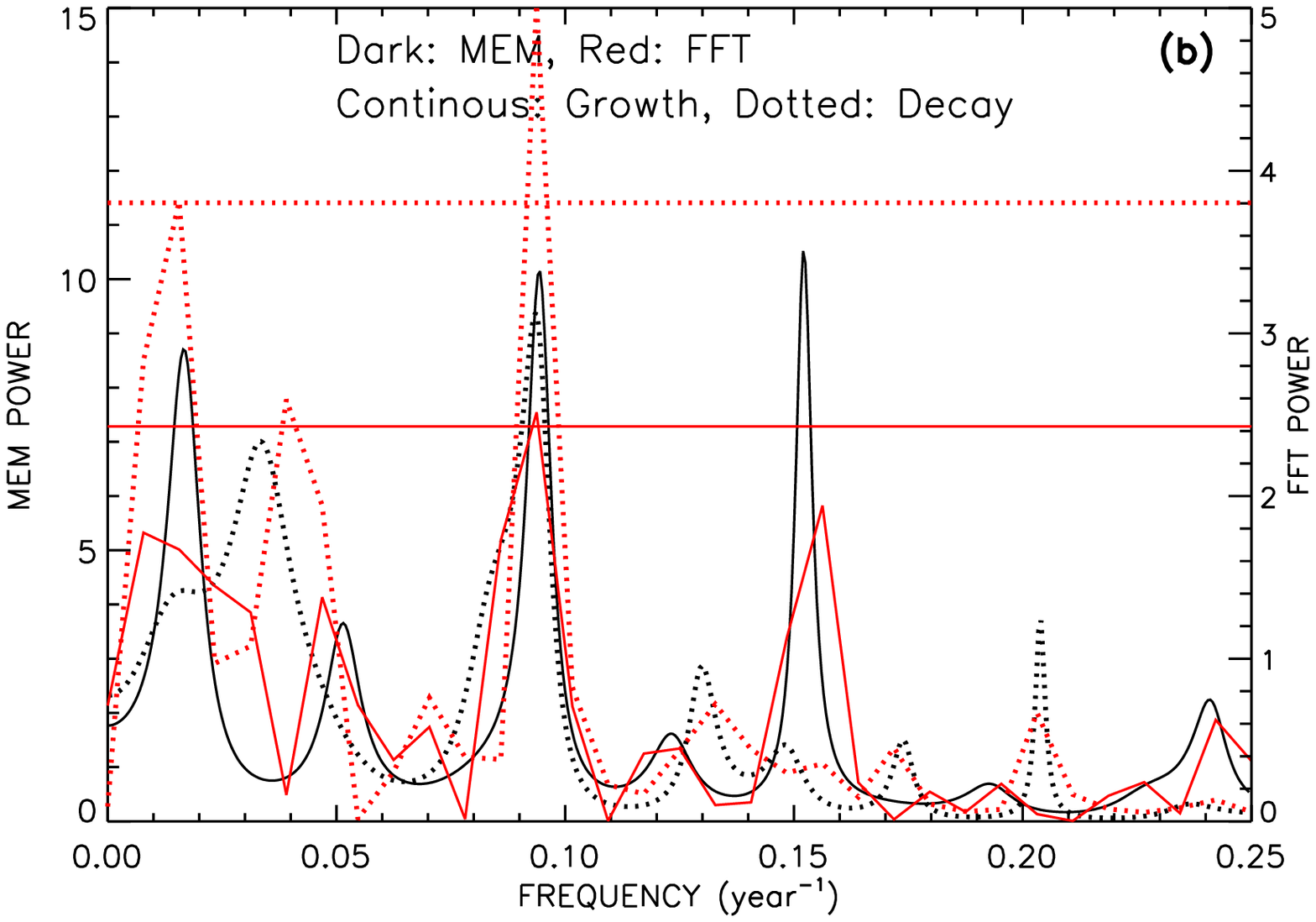}}
\caption{FFT  and MEM  spectra 
of $\overline {\%G}$ and $\overline {\%D}$  determined from 
the  data in $0^\circ - 10^\circ$ latitude intervals of  
(a) the northern hemisphere and (b) 
the southern hemisphere, during  4-year MTIs shown in  Figures~4 and 5.} 
\end{figure}

\begin{figure}
\centerline{\includegraphics[width=\textwidth]{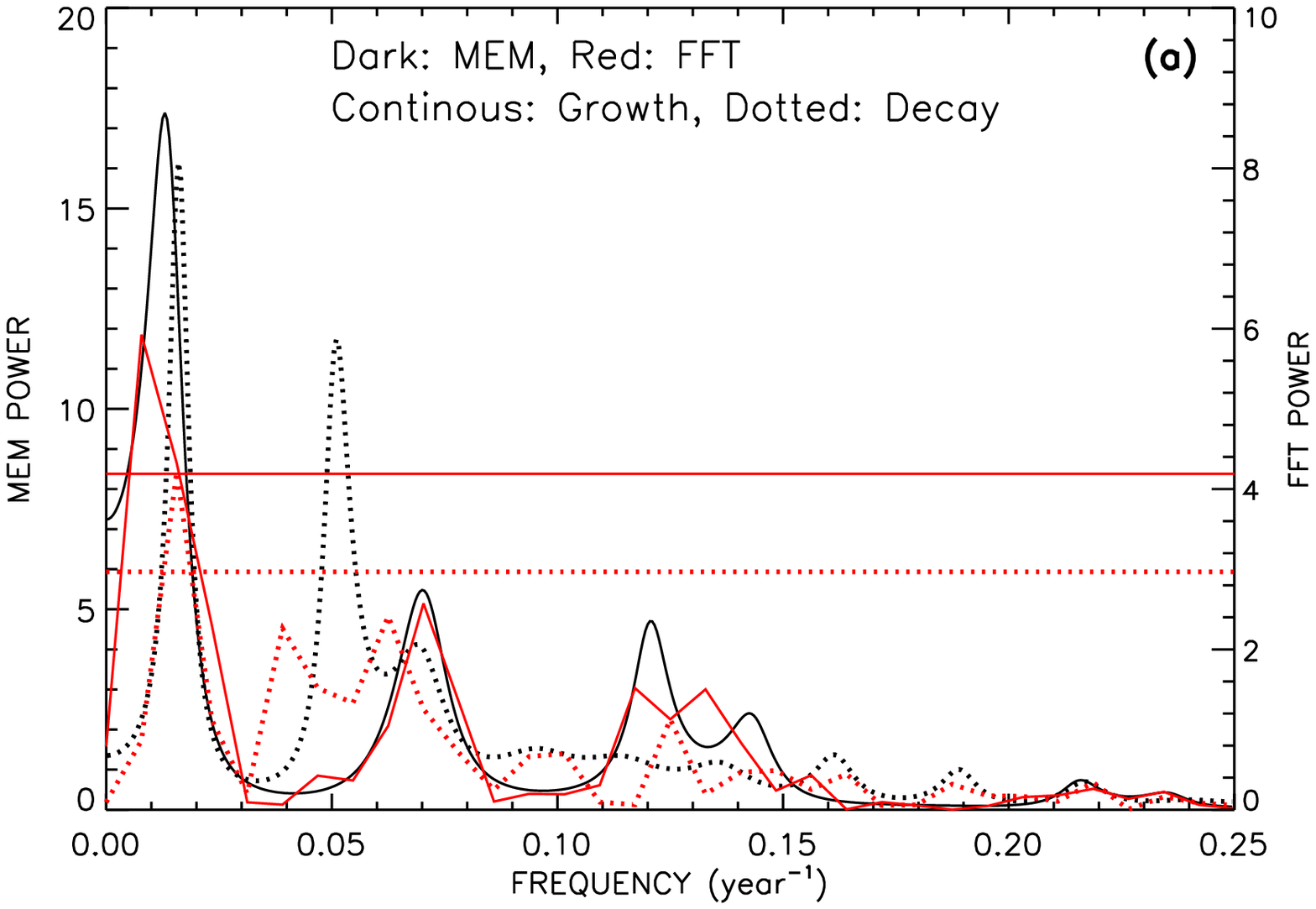}}
\centerline{\includegraphics[width=\textwidth]{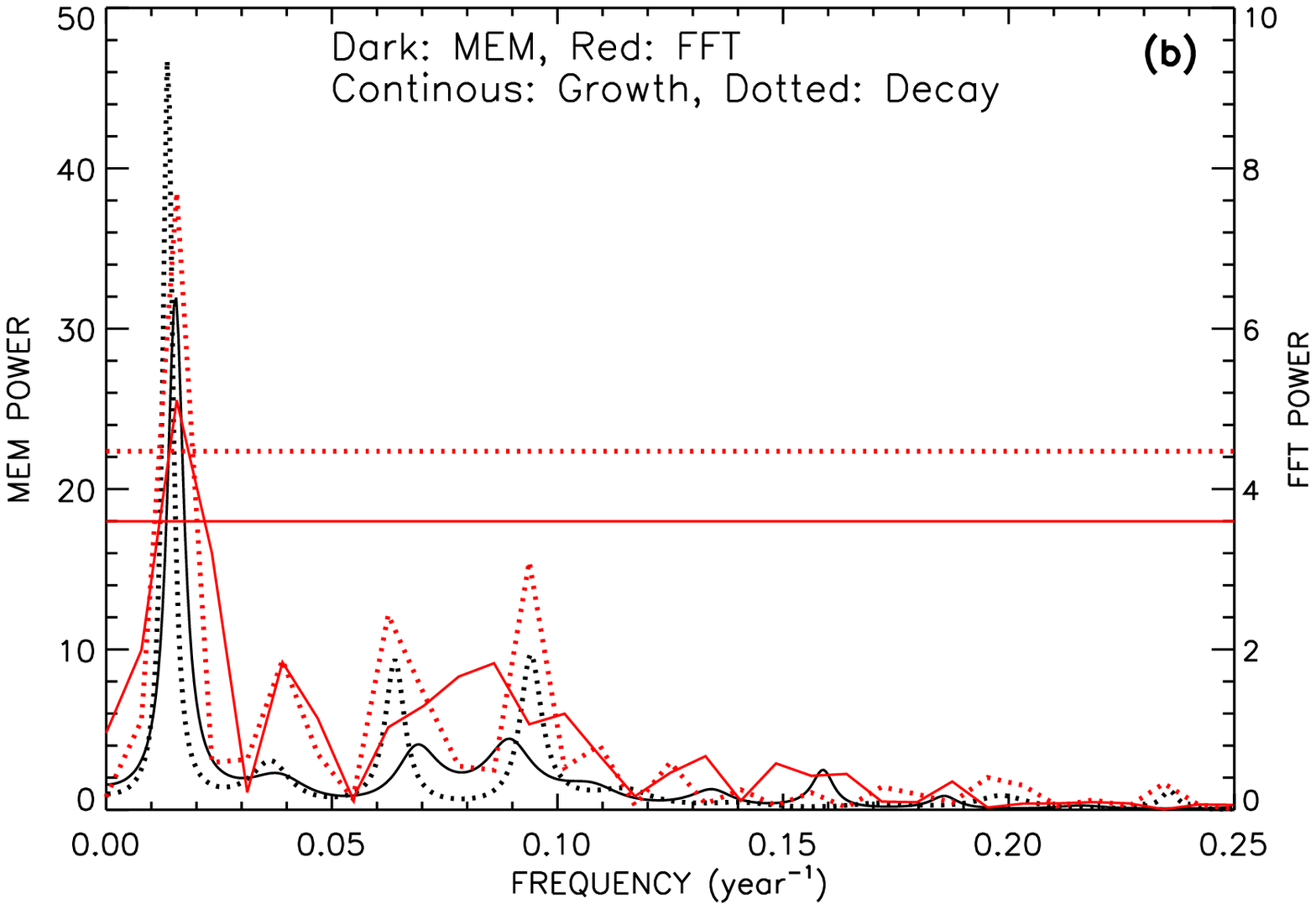}}
\caption{The same as Figure~10, but for the data in
  $10^\circ - 20^\circ$ 
latitude intervals.}
\end{figure}

\begin{figure}
\centerline{\includegraphics[width=\textwidth]{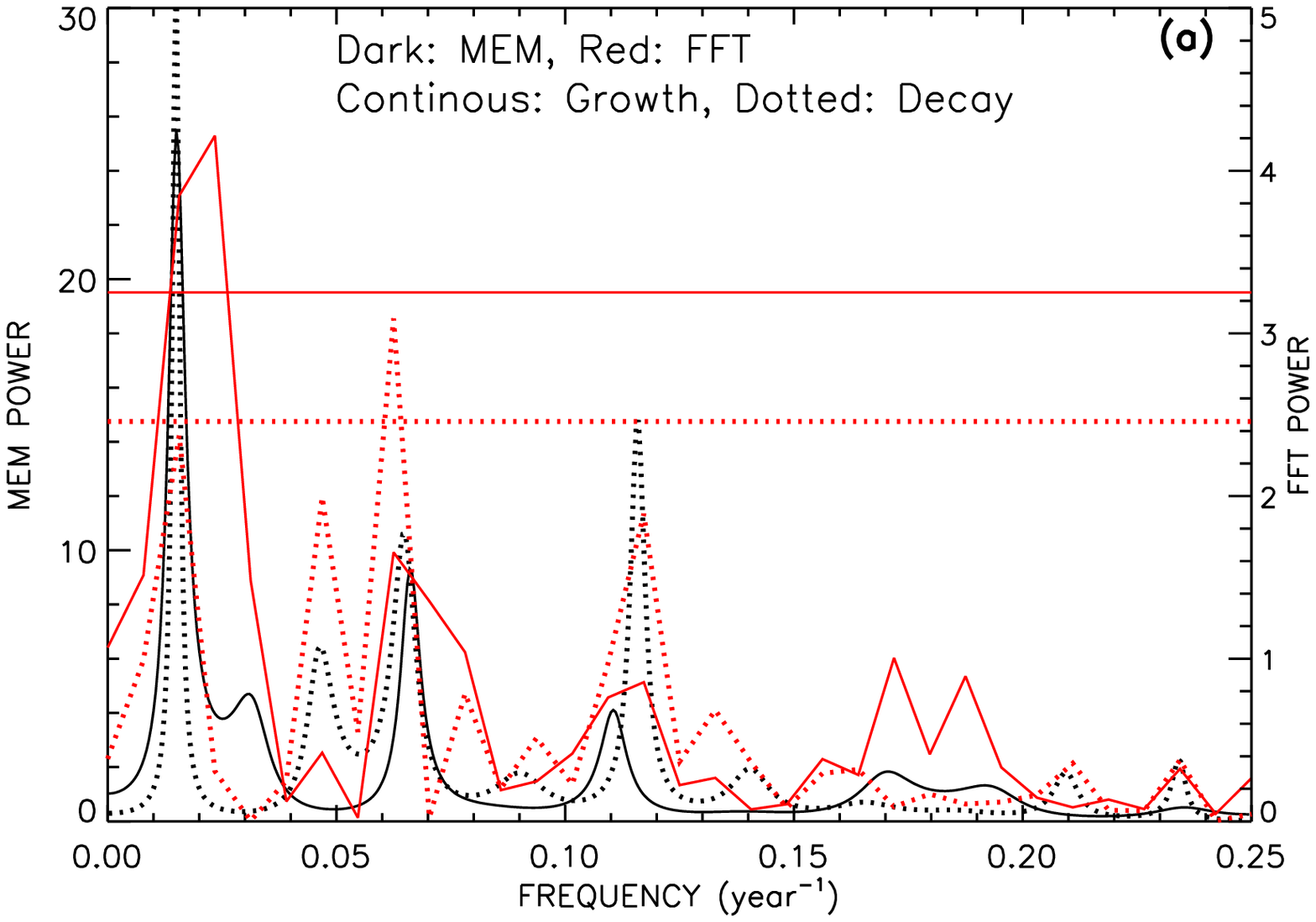}}
\centerline{\includegraphics[width=\textwidth]{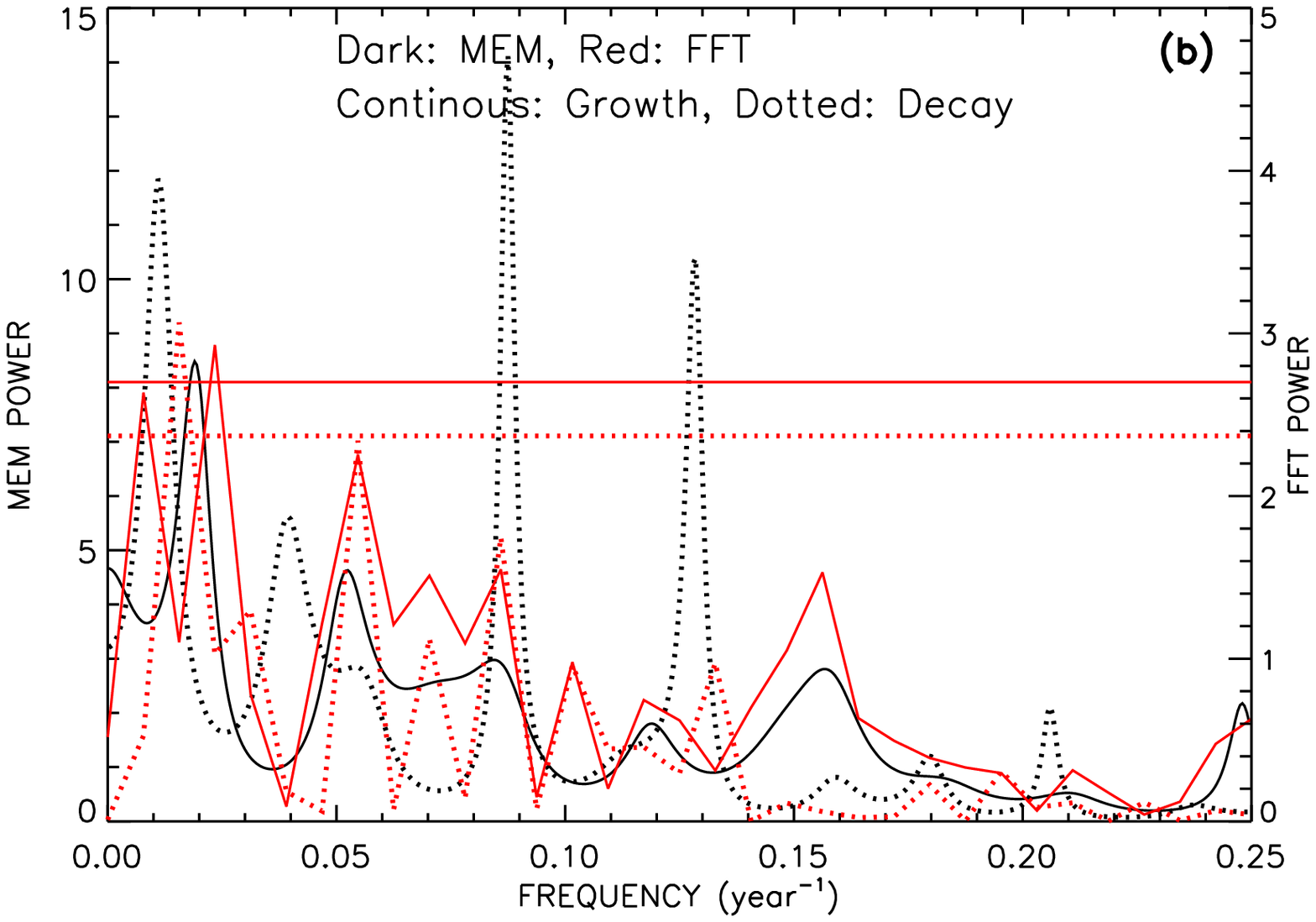}}
\caption{The same as Figure~10, but for the data in 
 $20^\circ - 30^\circ$ 
latitude intervals.} 
\end{figure}

\begin{figure}
\centerline{\includegraphics[width=11.5cm]{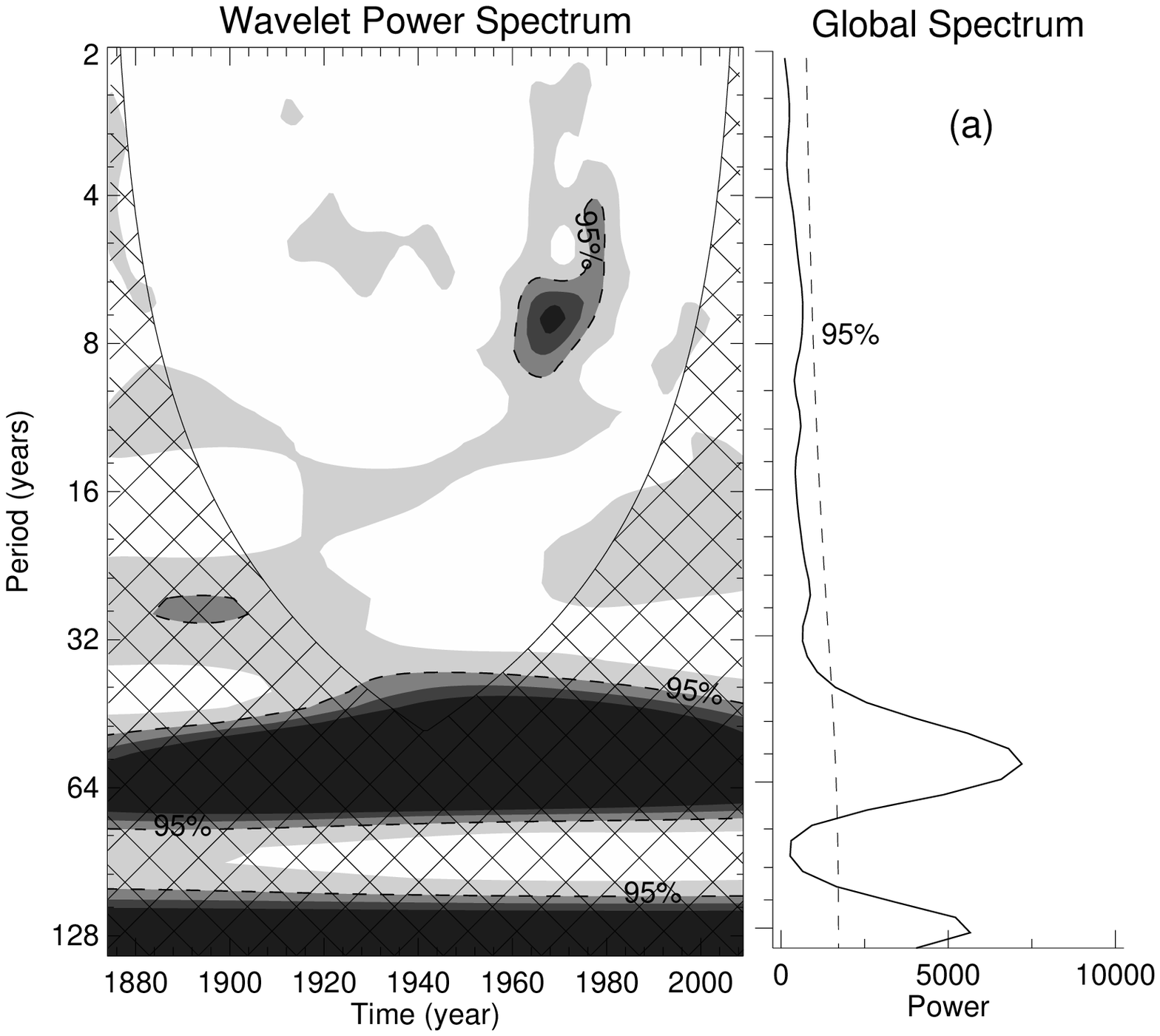}}
\centerline{\includegraphics[width=11.5cm]{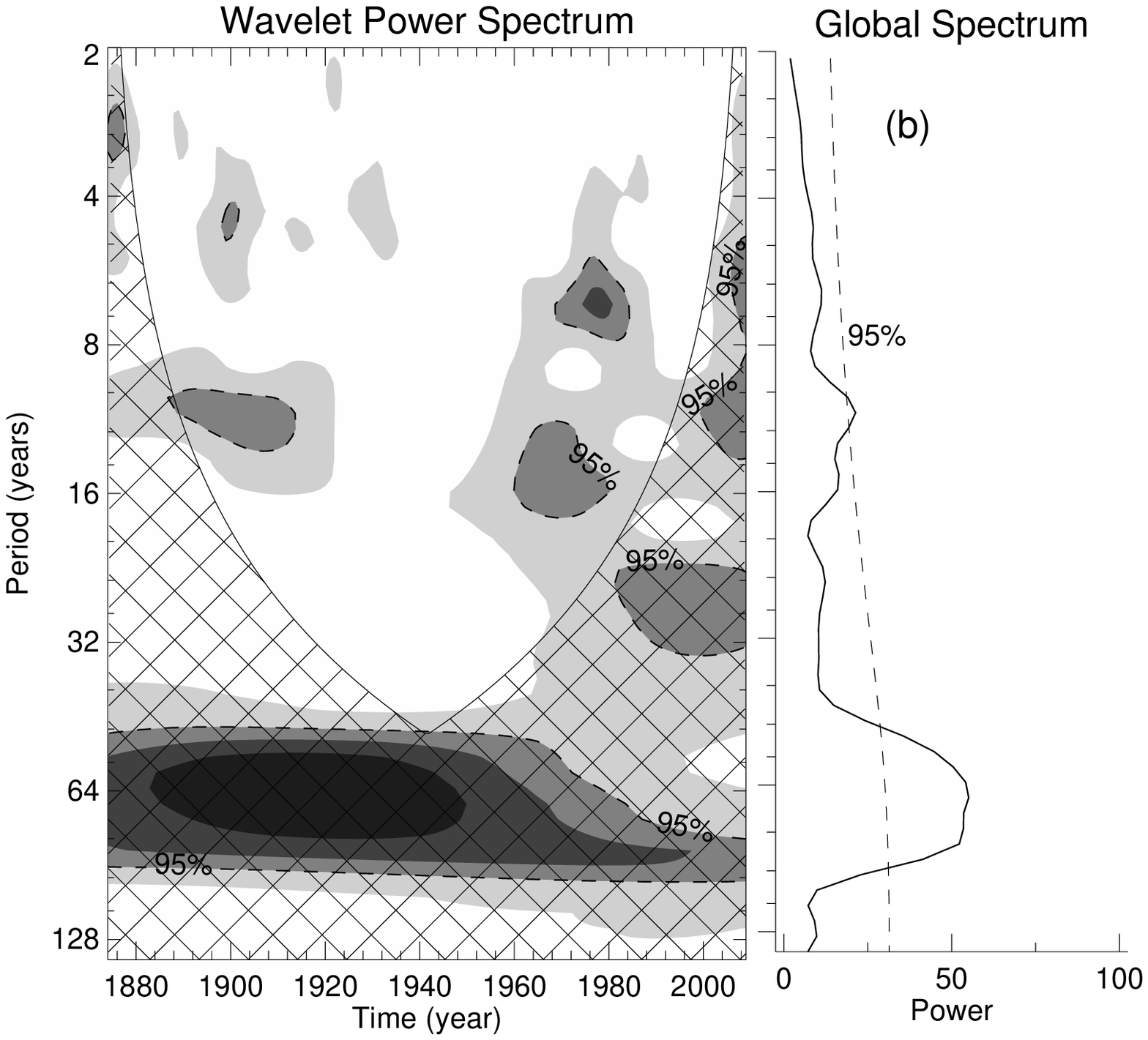}}
\caption{Wavelet power spectra and  global spectra of the corrected
 yearly time series of  $\overline {\%G}$ (upper panel) and
 $\overline {\%D}$ (lower panel) 
  shown in Figure~2.   The wavelet spectra are normalized by the variances
 of the corresponding time series. The shading are the normalized variances of
 1.0, 3.0, 4.5, and 6.0, respectively. The dashed curve represents the 95\%
 confidence level, deduced by assuming a white noise process. The cross-hatched regions indicate the "cone of influence", where edge effects become important 
(Torrence and  Compo, 1998).}   
\end{figure}

 As can be seen in Figure~2,  
the minimum and the maximum  values of $\overline {\%G}$ 
are     $\approx$52 and  $\approx$183, respectively, and the corresponding   
  values of $\overline {\%D}$ 
are   $\approx$21 and  $\approx$44, respectively (we have given here the 
maximum and the minimum values in the corrected annual time series).
The average value (over the period 1874\,--\,2009)
of the growth rate  
 is about 70\% higher than that of the  decay rate.
Figure~6 shows the   plots of the yearly mean values 
(correspond to the corrected annual time series)  
 of  $\overline {\%G}$ and $\overline {\%D}$ {\it versus} the 
year of the solar Cycles~12\,--\,23  
 (the point at year 9 belongs to the beginning year of the 
current Cycle~24 and  the data are available only 
for the last four years of Cycle~11).  To recognize and  compare between the variations during 
individual cycles,  
 the odd- and   even-numbered cycles data  are plotted with 
different colors,  and  different symbols are used for  different pairs 
of the odd- and even-numbered cycles.
In Figure~6(a) the overall spread in the data points of  $\overline {\%G}$ is 
 very  large.
 Thus, the
variations  during different solar cycles
highly differ (deviations from the mean 
variation are large  in the cases of Cycles~15 and 20), 
 suggesting that there is no  statistically  
significant 11-year periodicity  in   $\overline {\%G}$.   
In the case of Cycle~20, in which the deviation  is largest, there is a high
anticorrelation between  $\overline {\%G}$ and  amount of  activity.  
 As can be seen Figure~6(b), 
the mean pattern of $\overline {\%D}$ 
 suggests that
 $\overline {\%D}$ 
 decreased from the beginnings of the cycles, reached minimum
 at 1\,--\,2  years after the maximum epochs, and then increased up to 
near the   ends of the cycles. However, 
the spread in the data points of 
 $\overline {\%D}$  is large before the epoch -2 and after the epoch 
2. 
 That is, there is a large ambiguity in the mean pattern.
Thus, the  11-year cycle is not quite evident 
even in $\overline {\%D}$.  
The variations in $\overline {\%D}$  during a large numbers of the
 odd-numbered cycles seem to be little more  
closer to the mean variation than those of even cycles. 
 However,  the deviation from the mean variation is very large in the case of 
Cycle~21. 
In the case of Cycle~23,
which is an 
anomalous cycle 
in the sense that the cycle pair 22,23 violated the Gnevyshev and Ohl rule
(according to this rule
 a preceding even-numbered cycle is weaker than its following 
odd-numbered cycle), 
$\overline {\%G}$ decreased 
substantially  
from starting to  near 
the beginning of the 
 declining phase  
and then  
remained in the same low level 
for   a long time,     
near the end it is smallest 
in the last about 100 years. In fact, it  is close to the corresponding
 small value in the declining phase of Cycle~11.  Interestingly, the length of the
 declining phase of  Cycle~11 is also substantially long 
(it is longer than those of Cycles 12\,--\,22).
In  Cycle~23 the  variation in $\overline {\%D}$ 
is close  (less deviation)  to the mean variation 
from the beginning year to 
 $\approx$ 11th year, but it 
   strongly increased 
 during the extended part 
of the declining phase  (beyond the
 duration of the declining phase of a normal cycle) 
of this cycle
 and it is largest in the beginning of the 
current Cycle~24. 
These unusual properties of $\overline {\%G}$  and
  $\overline {\%D}$ in Cycle~23 may be related to the process which 
 caused the very  
 long decline phase  of this cycle  with the 
 unusually deep and prolonged current minimum.

As can be seen Figures~2 and 3, 
 the  average values of $\overline {\%G}$ 
 over the periods 1910\,--\,1940 
and 1960\,--\,1980 seem to be larger than the corresponding 
average values over the periods 1874\,--\,1910,
 1940\,--\,1960, and 1980\,--\,2008. Moreover, although there are 
fluctuations of several time scales, one can see the  following trends:
   an increase in $\overline {\%G}$ 
from 1874 to 1920,    a decrease  from 1920 to 1940 and then 
an increase up to 
1965,  and  again a  decrease up to 2008. 
This  pattern 
 suggests the existence of
a $\approx$60-year cycle 
in $\overline {\%G}$. 
The  average values of $\overline {\%D}$ 
 over the periods 1900\,--\,1920 
and 1960\,--\,1990 seem to be somewhat larger than the corresponding 
averages values over the periods 1874\,--\,1900,
 192\,--\,1960, and 1960\,--\,2008.  
Moreno-Insertis and  V\'azquez (1988) and  Mart\'{\i}nez Pillet,
Moreno-Insertis, and V\'azquez (1993) found the evidence for 
a significant variations in the decay rates of the spot groups of
  Cycles~12 through 16. We also see 
such an evidence in the variation of $\overline {\%D}$. In fact,
one can see the following trends: an increase in  $\overline {\%D}$
   from 1874 to 1900
 and then a decrease up to 
1945,  again an  increase up to 1980 and then  a decrease.
This  pattern  suggests  the 
existence of a 60\,--\,80-year cycle in $\overline {\%D}$.  
 On the other hand, there is  only a weak correlation (25\%\,--\,40\%) between 
 $\overline {\%G}$ and $\overline {\%D}$, indicating   that there may be 
 differences in 
 the  long-term trends of the 
 $\overline {\%G}$ and $\overline {\%D}$, which cannot be detected 
in the present analysis due to inadequate data. Overall, we find  that 
$\overline {\%G}$
  varies by about 35\% on a 60\,--\,80 year time scale,
 whereas $\overline {\%D}$ seems to vary by about 13\% on nearly 
the same time scale.

Figure~7 shows the cycle-to-cycle modulations in  $\overline {\%G}$ and 
 $\overline {\%D}$  determined by averaging the corrected annual  values
(shown in Figure~2)  over the duration of each of 
the Cycles~12\,--\,23. In the same figure the variation in the 
amplitudes  of the cycles ($R_{\rm M}$, the largest smoothed monthly mean 
 sunspot numbers, taken from the website, 
{\tt ftp://ftp.ngdc.noaa.gov/STP/SOLAR$\_$DATA/SUNSPOT\break $\_$NUMBERS}), is also shown.
 From this figure one can see clearly  the  long-term   
variations described above, $i.e.$, a cycle of  about 55 years
 in $\overline {\%G}$ 
with minimum epoch at Cycle~18 and maximum epochs  at Cycle~15 and 20, 
  and a cycle of about 65 years 
in $\overline {\%D}$  with  minimum epoch around Cycles~18\,--\,19 and 
 maximum epochs at Cycles~15 and 21. 
The correlation between  $\overline {\%G}$ and $\overline {\%D}$ from
the  12 pairs of 
 data points is only 78\%.
The 10-year difference between  the periods of these 
cycles might have caused a
 few year phase catastrophe and responsible for the weak correlation.   
There is no  correlation (values are very small, 6\%\,--\,12\%) either 
between 
  $\overline {\%G}$ and $R_{\rm M}$ or between 
$\overline {\%D}$ and $R_{\rm M}$.
In addition, 
 $\overline {\%G}$ is considerably large in the declining phases of
 some solar cycles  and $\overline {\%D}$ is considerably
 large in the rising phases of some cycles (see Figure~6). 
Therefore, the relation between the sunspot activity and  the evolution rates
 of the   spot groups is not clear.   
On the other hand, 
the different types of cycles of   the sub-century and the 
century time scales  may be dominant in  indices of different types of activity 
 (Komitov $et\ al.$, 2010). 
   Subsurface  and 
surface flows may be responsible for the growth and the decay of
 the spot groups (active regions in general).  
 It may be interesting to note 
that the correlation between the powerful flare events 
and sunspot number is weak and a significant 
50\,--\,60-year periodicity seems to be highly pronounced  in the powerful
 flare-related data
 (Komitov  $et\ al.$, 2010). Evolution of sunspot groups 
may be related to the cancelation/enhancement of the subsurface/surface 
 magnetic flux  due to the  
reconnection process, which seems to be the main mechanism of 
solar flare 
activity.   Hence,  the long-term variations   in  $\overline {\%G}$ 
and $\overline {\%D}$  may be related to the corresponding long-term 
variations in the powerful flare events.

Although over the 12 cycles  the values of the  correlations 
between  $\overline {\%G}$ and $R_{\rm M}$,
 and between $\overline {\%D}$ and  $R_{\rm M}$, are small, in Figure~7 
there is 
a suggestion that  the
  increasing/decreasing 
 trends of both  $\overline {\%G}$ and  $\overline {\%D}$ during 
  3\,--\,4 cycles
are related to   
the mean  $R_{\rm M}$ of the corresponding cycles, which is lower/greater than 
the 
mean $R_{\rm M}$ of the 12 cycles. In addition,  
 the $\approx$55\,--\,60-year
 cycle looks to be  superposed (particularly in the case of $\overline {\%G}$)
 on a much stronger cycle of $\approx$ 120-year (with minima
at Cycles~12 and 23). In this context it may be worth to note that 
 the length of the current Gleissberg cycle is also 
  substantially long, $113 \pm 5$ years 
(Javaraiah, Bertello, and Ulrich, 2005).   
These  trends in $\overline {\%G}$  and $\overline {\%D}$ 
 indicate that both these parameters will slightly  increase 
during the next 2\,--\,3 cycles and 
the behavior may be similar to the corresponding trends
of $\overline {\%G}$  and $\overline {\%D}$ 
 during Cycle 12\,--\,15. 
Since, Cycles~12\,--\,15 are on the average weak  cycles,
 hence the next few cycles may be
 weak cycles.  
This inference is also consistent with the following suggestions:  
 (i) the violation of the G-O rule by cycle pair 22,23 might be
  followed by a few weak cycles (Javaraiah, 2005); (ii) 
now the level of activity may correspond to the declining phase of the current 
 Gleissberg cycle (Javaraiah, Bertello, and Ulrich, 2005); and (iii)
 a small amplitude predicted for solar Cycle~24 (Javaraiah, 2008 and 
references therein).

 We find the  high correlations 69\% and 74\%   between the northern 
and the southern hemisphere 
$\overline {\%G}$ and $\overline {\%D}$ shown in Figure~3, respectively,
 suggesting that
 the  long-term 
variations in $\overline {\%G}$ and $\overline {\%D}$ are closely symmetric
about the Sun's equator. However, 
as can be seen in Figure~3, 
 the phases of the variations of $\overline {\%G}$ 
in the northern and the southern hemispheres   
 seem to be opposite 
 during each of the Cycles~17\,--\,19. In fact, we obtained -74\% correlation 
  from 20 pairs of data points of $\overline {\%G}$ 
correspond to the time number interval 65\,--\,84, $i.e.$, between around
 the maximum epochs 
 of Cycles~17 and 19. On the other hand, in the recent four cycles 
there is no  north-south asymmetry in $\overline {\%G}$, indicating that   
a  high coherence in the variations of $\overline {\%G}$ of
 the northern and the  
southern hemispheres is associated with reduced
amplitude of a 33\,--\,44-year cycle (may be due to equatorial crossing of 
the magnetic flux in the convection zone).
Thus, the strength of a 33\,--\,44-year cycle
 in the  solar activity may be related to 
the strength of the  north-south 
asymmetries in  $\overline {\%G}$ during  the 11-year solar cycles 
that  comprise  the 33\,--\,44-year cycle in activity.
There is no north-south asymmetry  in 
 $\overline {\%D}$
throughout the period 1874\,--\,2009, 
except during Cycle~17.

Figures~8\,--\,12  show the FFT and  the MEM power spectra of 
  the variations in $\overline {\%G}$ and 
$\overline {\%D}$ shown in Figures~2\,--\,4 (the last few years data 
have  not been used in case of all the time series of the 4-year MTIs, 
because their inclusion found to distort  
 the  low  frequency sides of the corresponding MEM spectra). It should be noted that as 
the size of the time-interval increased  the peaks at  high frequencies 
are  washed-out. Therefore, 
only low frequency sides 
of the spectra are shown in Figures~9\,--\,12. However, the 95\% significant levels of
the peaks in the FFT  spectra  are determined by considering only the power
 at these low frequencies.   
(Note:  
 the spectra which correspond to the $20^\circ - 30^\circ$  latitude interval 
 are less reliable due the 
scarcity of the data in many 4-year  time-intervals.)  Most of these  
spectra show the existence of a  60\,--\,80-year periodicity in both   
$\overline {\%G}$ and $\overline {\%D}$, 
with the corresponding peaks in the 
  FFT spectra are about 95\% 
confidence level. (The MEM spectra of the yearly data of 
 $\overline {\%G}$ and $\overline {\%D}$ 
yield the values 59-year and 
71-year, respectively, for the 60\,--\,80-year periodicity.) In addition, 
the following  peaks  
   at  frequencies which are close to the 
  solar cycle frequency are significant on $\ge$ 95\% confidence level: 
 (i) the peak at frequency $\approx$ 0.1 year$^{-1}$ in 
the FFT spectra of both $\overline {\%G}$ and $\overline {\%D}$ 
of $0^\circ - 10^\circ$ latitude interval of 
the southern hemisphere (see Figure~10(b)); and 
  (ii) the peak at frequency
 $\approx$ 0.055 
year$^{-1}$ in  
the FFT spectrum of $\overline {\%D}$ of $20^\circ - 30^\circ$ latitude interval of
the northern hemisphere (see Figure~12(a)).
These results suggest 
that the  periodicities which are close to the solar cycle period 
are time and latitude
 dependent. Figure~13 shows the wavelet spectra 
of the corrected time series shown in Figure~2.
 The lengths of the time series 
are inadequate to resolve the temporal dependency of 60\,--\,80-year
 periodicity. 
The periodicities of $\overline {\%D}$ which are close to  the solar-cycle
 period seem to have occurred 
during the period 1900\,--\,1920 and around the period 1975\,--\,1995.
 A 7\,--\,8-year periodicity  
in $\overline {\%G}$ and also in  $\overline {\% D}$  seems 
 to occur  during the period 
1960\,--\,1980. A $\approx$ 5-year periodicity
in $\overline {\%D}$  occurred around 1900.

 Hathaway and Chowdhury (2008) noted that the solar cycle variation in 
the decay rates of the spot groups is merely coming from 
the latitude-time distribution of the spot groups.  The existence of a 90-year 
 cycle is also 
found in the  widths of the Butterfly diagram (Pulkkinen $et\ al.$, 1999). 
Therefore, one can suspect that  
 the long-term ($\approx$ 60-year) periodicity found here in 
$\overline {\% G}$ and $\overline {\%D}$ may also be artifacts of the long-term 
variations in the width of the  Butterfly wings.  
On the other hand,  in Figures~4 and 5  we can see  that 
all the above said properties 
in the variations of $\overline {\%G}$ and $\overline {\%D}$ 
 are present in each of the  $10^\circ$
 latitude 
intervals, though  the patterns  are  not as well defined
 as those of $\overline {\%G}$ and $\overline {\%D}$   in a whole hemisphere.
However, a large contribution 
to  the  60\,--\,80-year variations in
 $\overline {\%G}$ and $\overline {\%D}$ of a whole hemisphere
is came   from  the corresponding variations in the 
$10^\circ - 20^\circ$ latitude interval (see Figure~11).

\section{Discussion and Conclusions}
The following conclusions can be drawn from the above
 analysis:
\begin{enumerate}
\item The minimum and the maximum values of the annual mean growth rates
of the spot groups  are 
  $\approx$52\%  day$^{-1}$ and 
$\approx$183\%  day$^{-1}$, respectively, whereas the corresponding values of 
the annual mean decay rates are 
  $\approx$21\% day$^{-1}$ and  $\approx$44\%  day$^{-1}$, respectively.
The average value (over the period 1874\,--\,2009)
of the daily growth rate  
 is about 70\% more than that of  the decay
rate.
\item The growth and the decay rates vary by about 35\% and 13\%, respectively, on a 60-year
 time scale.
\item A $\approx$ 11-year periodicity in the growth and the decay rates 
 is  found to be highly latitude and time 
dependent and seems to exist  in both the growth and the decay rates of
 the spot groups mainly in  the $0^\circ - 10^\circ$ latitude interval
 of the southern hemisphere.
\item  From the beginning of Cycle~23  
the growth rate is substantially decreased and   near the end
(2007\,--\,2008) 
the growth  rate is lowest in the past about 100 years. 
In the extended part (beyond the length of the declining part of a 
normal cycle) of the
 declining phase of this cycle, the decay 
rate steeply increased  and it is largest in the
 beginning of the current Cycle~24. 
These unusual properties of the growth and the decay rates  during Cycle~23
  may be related to the  very
  long declining phase of this cycle  with the 
 unusually  deep and prolonged current minimum.
\item The strength of the known   approximate 33\,--\,44 year modulation
in  solar activity seems to be related to  the strength of the
 north-south asymmetry in the growth rate during the 
11-year solar cycles that comprise the 33\,--\,44 year cycle in activity. 
\item  Decreasing and  increasing trends in the growth and the decay 
rates indicate that the next 
 2\,--\,3 solar cycle will be relatively weak. 
\end{enumerate}

The conclusions~(1) and (2) above,  $i.e.$, the lower decay rate and its slow
evolution 
 support the idea of a long-term persistence 
of the solar activity. This is consistent with an hypothesis of 
 some class of turbulent  
dynamo models (Dikpati and Gilman, 2006), although the present analysis 
(conclusion 6, above) does not support  the large amplitude  for 
 the current Cycle~24 that is predicted in these models.  

 The existence of a quasi-periodic long-term variation 
(Gleissberg cycle) in solar activity is well known. 
However, 
a  wide range of values (50\,--\,130 years)  are suggested   for 
Gleissberg cycle. 
 Among them    
the existence of 
  60\,--\,80-year periodicity is  
 frequently 
reported (Komitov  $et\ al.$, 2010).  
  The 60\,--\,80-year  cycles of the growth and decay rates
may be
related to the corresponding cycles in the powerful flare events,
rather than to that of the sunspot activity 
(see previous section).  

One can expect that  convection has a major role in the
 day-to-day 
fluctuations of the areas of  spot groups. 
The growth and decay of an active region may have an  
association with upward  flow and the reverse down flow
 of the convection, respectively (see also Komm, Howe, and Hill, 2009). 
The long-term variations in the day-to-day fluctuations in the 
areas of the spot groups largely represent the corresponding long-term 
variations in the convective flows.
The long-term variation in the convection may be related  to 
the   corresponding 
long period global oscillation of the Sun.

\acknowledgements{The author thanks the anonymous referee for useful comments 
and suggestions. Wavelet software was provided by Ch. Torrence and G.P. Compo, 
and is available 
at {\tt URL: http//poas.colorado.edu/reserch/wave\break lets/}. The MEM FORTRAN code
was provided to us by Dr. A.V. Raveendran.}

\end{article}
\end{document}